\documentclass[aps,showpacs,twocolumn,nofootinbib]{revtex4}
\usepackage{graphicx}
\usepackage{dcolumn}
\usepackage{bm}
\newcommand{\beq}{\begin{equation}}
\newcommand{\eeq}{\end{equation}}
\newcommand{\beqa}{\begin{eqnarray}}
\newcommand{\eeqa}{\end{eqnarray}}

\newcommand{\pla}{\text{Pl}}

\def\beq{\begin{equation}}
\def\eeq{\end{equation}}
\def\bea{\begin{eqnarray}}
\def\eea{\end{eqnarray}}
\newcommand{\be}{\begin{equation}}

\newcommand{\ee}{\end{equation}}
\def\beqa{\begin{eqnarray}}
\def\eeqa{\end{eqnarray}}
\def\beq{\begin{equation}}
\def\eeq{\end{equation}}

\let\gam=w

\renewcommand{\epsilon}{\varepsilon}

\def\lta{~\mbox{\raisebox{-.6ex}{$\stackrel{<}{\sim}$}}~}
\def\gta{~\mbox{\raisebox{-.6ex}{$\stackrel{>}{\sim}$}}~}

\newcommand{\tev}{\text{TeV}}
\newcommand{\gev}{\text{GeV}}

\begin{document}


\title{The Vacuum Frame}

\author{Jose A. R. Cembranos}
\email{cembra@ucm.es}

\affiliation{Departamento de F\'\i sica Te\'orica and IPARCOS, Facultad de Ciencias F\'\i sicas,
Universidad Complutense de Madrid, Plaza de Ciencias, n 1, 28040 Madrid, Spain}

\begin{abstract}
One of the most fundamental questions in cosmology is if dark energy is related just to a constant or it is something more complex. In this work, we call the attention to the fact that, under very general conditions, dark energy can be identified with a cosmological constant. Indeed,
this fact defines what we call {\it Vacuum Frame}. In general, this frame does not coincide with the
Jordan or Einstein frame, defined by the invariant character of particle masses or the Newton constant, 
respectively. We illustrate this question by the introduction of a particular scalar-tensor model
where the different hierarchies among these energy scales are dynamically generated. 
\end{abstract}


\maketitle

\section{Introduction}

The present expansion of the Universe is accelerated due to a non-standard component called Dark Energy (DE). One of the most basic questions in cosmology is to characterize its fundamental nature. The most simple model of DE is just a cosmological constant $\breve{\Lambda}$. Indeed, it determines (part of) the standard model of cosmology, named $\Lambda$-CDM. However, the uncertainties in this respect are important and many observational efforts are and will be dedicated to elucidate this question.

In this work, we would like to point out that the constant nature of DE can be imposed in the majority of cases by taking advantage of a conformal transformation that determines the measurement pattern for energy scales. We have the freedom of choosing such a pattern in any point of our spacetime. If we choose it precisely proportional to the DE, such DE will be constant. This is what defines what we name {\it Vacuum Frame} (VF). We can summarize the two basic necessary conditions in order to can properly define the VF, at least, in a given spacetime domain:
\begin{itemize}
\item[{1.-}] DE cannot be conformal invariant. This condition is obviously needed in order to modify the DE profile with a conformal transformation.
\item[{2.-}] DE cannot be zero in any point of such domain. This fact is also necessary to can define an energy pattern within the entire domain. 
\end{itemize}
Working within the VF provides an interesting different approach to the DE problem. We will illustrate it with a particular scalar-tensor theory (STT). This type of models refers to any gravitational theory, whose interaction is mediated not only by the standard spin-2 graviton, but also by a spin-0 scalar. Under this definition is possible to classify many different extensions of General Relativity (GR) \cite{stgen,Kalara:1990ar}. The action of the model can be written in different forms
that are equivalent as a classical field theory. For example, it is usually written in the so called Jordan Frame (JF), in which the matter content is coupled to the metric in the minimal way, or in the Einstein Frame (EF), where the action for
the spin-2 graviton is standard. The metric tensors in these frames are related by a simple conformal transformation
\cite{stgen,Kalara:1990ar}:
\begin{equation}\label{jf_to_ef}
 g_{\mu\nu} = A^{2}(\varphi) g_{\mu\nu}^*\,,
\end{equation}
where $g_{\mu\nu}$ denotes the metric in the JF and $g_{\mu\nu}^*$, in the EF (we will use $*$ for EF quantities).
An interesting property of these theories is that physical scales are not absolute or constant. On the contrary,
they evolve due to the presence of the new spin-0 field: $\varphi$. In the EF, the Planck scale is constant ($M_{\pla}^*$) but
the scales associated with the particle content (that we will denote by $m^*$, in general) evolves proportional to the
conformal factor: $m^*=A(\varphi)\,m$. In the JF, the scales associated with the particle content are fixed (m) but not
the Planck scale: $M_{\pla}=M_{\pla}^*/A(\varphi)$.

\section{Vacuum Frame: an example}

In this work we are interested in written the model within the VF, where the
scale associated with the vacuum density is constant. We will denote the quantities associated with this frame with $\breve{}$.
Therefore, the action in the VF for the STT can be written as:
\begin{eqnarray}\label{actionVF}
  S &=&\int {d^4 x }\sqrt{-\breve{g}}\left[
  \frac{\breve{\varphi}^2}{2}\breve{R}
     -\frac{\breve{\omega}(\breve{\varphi})}{2}\,\breve{g}^{\mu\nu}
     \partial_{\mu}\breve{\varphi}\partial_{\nu}\breve{\varphi}
     -\breve{\Lambda}^4 \right]\nonumber\\
        &&   \qquad+ \breve{S}_{\text{SM}}[\breve{g}_{\mu\nu};\breve{\psi}; \breve{\varphi}]\,,
\end{eqnarray}
where $\psi$ denotes the different fields of the standard model (SM). We will be able to generate the Planck scale dynamically
by the vacuum expectation value (asymptotic value, strictly speaking) of the scalar $\breve{\varphi}_{\rm asy}=M_{\pla}^*$.

Therefore, the only explicit scale in the theory is the associated with the vacuum energy or cosmological constant:
$\breve{\Lambda}\simeq 2.3\times 10^{-3}$ eV. With respect to the matter sector, for simplicity, we will assume massless neutrinos and a conformal kinetic
term for the Higgs doublet ($\Phi$), although both assumptions can be trivially generalized. In this case, the only non-conformal
term in the matter action is the quadratic Higgs term. In order to generate the electroweak scale at the same time than
the Planck scale, we write this non-conformal term as:
\begin{eqnarray}\label{actionSMVF}
  \breve{S}_{\text{SM}}[\breve{g}_{\mu\nu};\breve{\psi}; \breve{\varphi}]&=&
  S_{\text{Conf}}[\breve{g}_{\mu\nu};\breve{\psi}]\nonumber\\
  &+&
  \int {d^4 x }\sqrt{-\breve{g}}\,
  \frac{\varsigma^2}{2} \breve{\Lambda}\,\breve{\varphi}\, \breve{\Phi}^\dagger \breve{\Phi}
  \,,
\end{eqnarray}
where $\varsigma$ is a dimensionless number, whose value needs to be fixed to
$\varsigma\simeq 0.037$. 
\begin{figure}[bt]
\begin{center}
\resizebox{8.0cm}{!} {\includegraphics{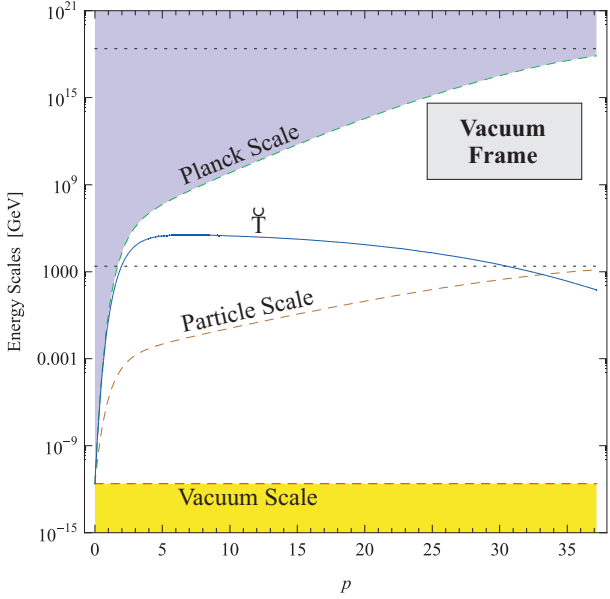}}
\caption{Evolution of the temperature, the Planck scale, vacuum scale, and the scale associated with the particle content inside the VF, where the vacuum density is constant by definition. The computation assumes the SM content and the first correction from QCD to the non-interacting thermal bath. We have chosen $\beta=80$.}
\label{Vacuum}
\end{center}
\end{figure}

\begin{figure}[bt]
\begin{center}
\resizebox{8.0cm}{!} {\includegraphics{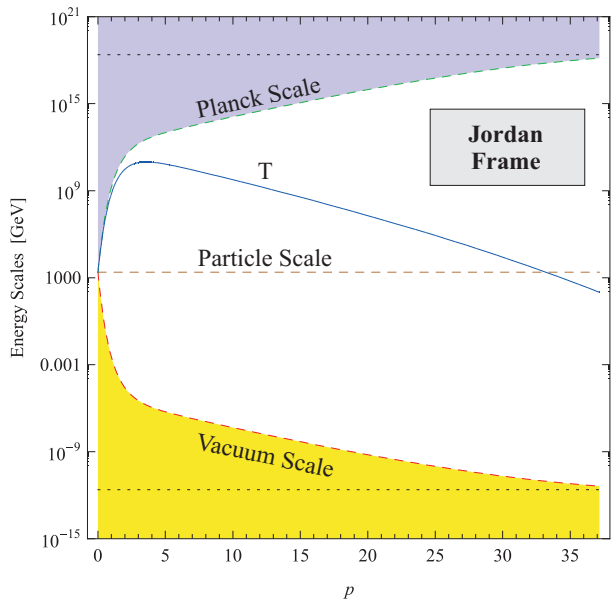}}
\caption{The same than Fig. \ref{Vacuum} but in the JF, where the scale associated with the particle content is constant
by definition.}
\label{Jordan}
\end{center}
\end{figure}

\begin{figure}[bt]
\begin{center}
\resizebox{8.0cm}{!} {\includegraphics{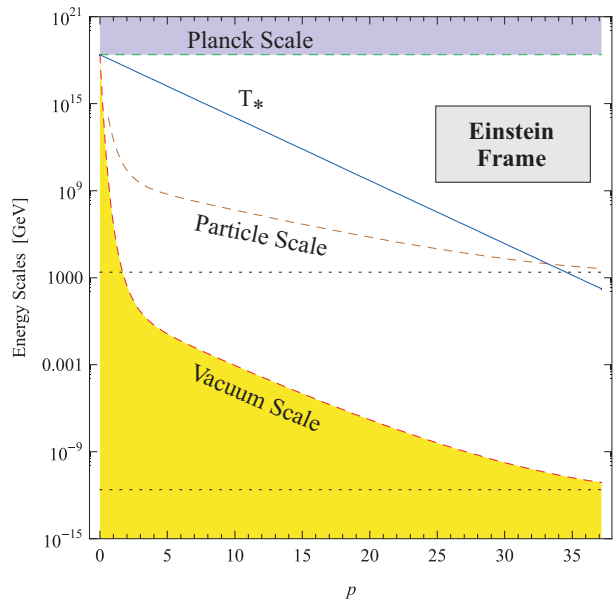}}
\caption{The same than Figs. \ref{Vacuum} and \ref{Jordan} but in the EF, where the Planck scale is constant.}
\label{Einstein}
\end{center}
\end{figure}

Although it is not evident from Eqs. (\ref{actionVF}) and (\ref{actionSMVF}), we will show that the new scalar mode
is coupled to the matter sector with a coupling $\alpha(\breve{\varphi}/\breve{\Lambda})$,
which is fixed by $\breve{\omega}(\breve{\varphi})$:
\begin{eqnarray}
\breve{\omega}(\breve{\varphi})&=&
\frac{1}{2}
\left(
\alpha^{-2}(\breve{\varphi}/\breve{\Lambda})
-12
\right)\,.
\label{wvf}
\end{eqnarray}
Therefore, to determine the STT we need to specify the function: $\alpha^2(x)$. For example, in
the classical Brans-Dicke theory: $\alpha^2(x)\equiv\alpha_0^2$ is constant. Unfortunately, in this case, an effectively
decoupled scalar field is necessary to be consistent with present constraints: $\alpha_0 \lta 10^{-2}$ \cite{Bertotti:2003rm}.
In this work we will focus on a different $\alpha^2(x)$ in order to show the potential of STT for 
dynamical generation of hierarchies. In particular we will focus on the function:
\begin{eqnarray}\label{coupling2}
\alpha^2(x)&=& \beta\left(b-\ln x\right)\,.
\label{alpha}
\end{eqnarray}
This type of couplings have been already proposed to describe the dilaton dynamics in non-perturbative string effective actions
\cite{polchy,dnord,dpol}. It has been studied in several works since it is the minimal version of the coupling which provides
an attractor toward GR \cite{dnord,dpol,couv}. In our case, the coupling given by Eq. (\ref{coupling2})
is even more interesting, since it is able to drive the value of the scalar field to the Planck scale without introducing any
fundamental hierarchy in the action, just by tuning $b=\ln[M_{\pla}^*/\breve{\Lambda}]\simeq 69$.
In addition, if $\beta$ is large enough, this scale can be generated dynamically. We take
$\beta = 80$ for Figs. \ref{Vacuum}, \ref{Jordan} and \ref{Einstein}. In particular, Fig. \ref{Vacuum}
shows that the present value of the Planck scale reaches its asymptotic limit:
$M_{\pla}^*\equiv (8\pi G_*)^{-1/2}\simeq 2.4 \times 10^{18}$ GeV.

It is simple to understand the dynamical generation of the Planck scale from Eq. (\ref{actionVF}). At the initial time in the VF,
any scale is expected to be of order $\breve{\Lambda}$. For the sake of concreteness, we assume
$\breve{\varphi}_{\rm ini}= \breve{\Lambda}$, which implies $\breve{M}_{\pla}(\varphi_{\rm ini})= \breve{\Lambda}$.
However, the effective coupling of the scalar mode is intense,
given by $\alpha(\breve{\varphi}_{\rm ini}/\breve{\Lambda}) = \sqrt{\beta\, b}$. 
It implies that $\breve{\varphi}$ is attracted efficiently toward its asymptotic value:
$\breve{\varphi}_{\rm asy}=\breve{\Lambda}\, e^{b}={M}_{\pla}^{*}$. In this limit,
$\breve{M}_{\pla}(\breve{\varphi}_{\rm asy})= \ M_{\pla}^{*}$ and $\alpha(\breve{\varphi}_{\rm asy}/\breve{\Lambda}) \rightarrow 0$,
which means that the scalar mode is decoupled and the model can be close enough to GR to satisfy precision constraints from
Post-Newtonian parameters (PPNs) or Big Bang Nucleosynthesis (BBN). For simplicity, we will not take into account an stage of inflation, but it should be taken into account that this stage would increase the
GR attraction. Therefore, the bounds found in this analysis can be qualified as conservative.

\section{Jordan Frame: The Standard Higgs Sector}

The linear coupling between the scalar mode and the Higgs doublet looks beyond the standard
STT approach, but it is not. The question it is that the above action is not written in either
the JF or the EF. The electroweak scale is determined by the vacuum expectation value of the Higgs
doublet, that depends on the scale: $\varsigma \sqrt{\breve{\Lambda}\,\breve{\varphi}}$.
We can define the JF as the frame in which the scale in front of ${\Phi}^\dagger {\Phi}$ is constant:
\begin{eqnarray}\label{actionSMJF}
  S_{\text{SM}}[g_{\mu\nu};\psi;\varphi]&=&S_{\text{SM}}[g_{\mu\nu};\psi]
  \nonumber\\
  & &
  \hspace{-2cm}
  =\; S_{\text{Conf}}[g_{\mu\nu};\psi]+
  \int {d^4 x }\sqrt{-g}\,
  \frac{(\varsigma\,\mu)^2}{2}\, \Phi^\dagger \Phi
  \,.
\end{eqnarray}
This frame is determined by the conformal transformation defined by the conformal factor,
$B^2(\breve{\varphi})=\breve{\varphi}/M_{\pla}^{*}$:
\begin{equation}\label{vf_to_jf}
 g_{\mu\nu}
 = (\breve{\varphi}/M_{\pla}^{*})\,\breve{g}_{\mu\nu}
 \;,\;\;\;\;
\breve{\varphi}/\breve{\Lambda}\equiv\varphi/\mu\,,
\end{equation}
which leads to the following action:
\begin{eqnarray}\label{actionJF}
  S &=&\int {d^4 x }\sqrt{-g}\left[
  \frac{\mu\,\varphi}{2}\,R
     -\frac{\mu\,\omega(\varphi)}{2\,\varphi}\,g^{\mu\nu}
     \partial_{\mu}\varphi \partial_{\nu}\varphi
     -\frac{{\mu}^6}{\varphi^2}
             \right]\nonumber\\
        &&   \qquad+ S_{\text{SM}}[g_{\mu\nu};\psi]\,.
\end{eqnarray}
We have introduced the scale: $\mu\equiv\sqrt{\breve{\Lambda}\,M_{\pla}^{*}}\simeq 2.4\;\tev$, and normalize the scalar mode in this frame such as $\varphi_{\rm ini}= \mu$. In this way:
\begin{eqnarray}
\omega(\varphi)&=&
\frac{1}{2}
\left(
\alpha^{-2}(\varphi/\mu)
-3
\right)\,,
\label{wjf}
\end{eqnarray}
and again, the action depends only on one scale: $\mu$, that is the scale associated with the particle sector, constant in this frame.
In this form, it is easy to recognize a standard STT in the JF, where the vacuum energy can be interpreted as a particular potential
for $\varphi$.

\section{Einstein Frame: Evolution}

For computation purposes, it is more convenient to write the action in the EF.
By Assuming a flat FRW metric: $d s^2 = -d t^2 + R_*^2(t)(d x^2+d y^2+d z^2)$,
and a perfect fluid for the matter content (characterized by its energy density, $\rho$ and its pressure $P$)
we can write the Friedmann equations in any frame. In particular, for the EF we can write: $\rho_*=A^4\,\rho$ and
$P_*=A^4\,P$, where $A$ is the conformal factor that defines the transformation from the JF. In our case:
\begin{eqnarray}
A^2&=&{M_{\pla}^{*}}^2/(\mu\,\varphi)=\exp(\beta \varphi_*^2),
\label{ein:A}
\end{eqnarray}
and the action in this frame reads:
\begin{eqnarray}\label{actionEF}
 S &=& \int d^4x \sqrt{-g_*}\frac{{M_{\pla}^{*}}^2}{2}\left[ R_*
        -2g_*^{\mu\nu} \partial_\mu\varphi_*\partial_\nu\varphi_*
        - 4V_*(\varphi_*)\right]\nonumber\\
  &&\qquad + S_{\text{SM}}[A^{2}(\varphi_*)g^*_{\mu\nu};\psi(\varphi_*,\psi_*)]\,.
\end{eqnarray}
where we are using a dimensionless scalar field $\varphi_*$ defined by
$\varphi=({M_{\pla}^{*}}^2/\mu)\,\exp(-\beta \varphi_*^2)$. Its potential reads:
%
\begin{eqnarray}
  2\,V_*(\varphi_*)&\equiv&\frac{\hat{V}_*(\varphi_*)}{{M_{\pla}^*}^2}
  ={M_{\pla}^*}^2\,e^{4\,\beta [\varphi_*^2-\varphi_{*\,\rm ini}^2]}\,.
  \label{ein:potential}
\end{eqnarray}
where $\varphi_{*\,\rm ini}^2\equiv b/\beta\sim 1$. In this frame, the non-conformal part of the
particle action depends also on $\varphi_*$:
\begin{eqnarray}\label{actionSMEF}
  S_{\text{SM}}[A^2(\varphi_*)\,g^*_{\mu\nu};\psi(\varphi_*,\psi_*)]&=&
  S_{\text{Conf}}[g^*_{\mu\nu};\psi_*]\nonumber\\
  & &
  \hspace{-4.4cm}
  +\,\int {d^4 x }\sqrt{-{g_*}}\,
  \frac{(\varsigma\,M_{\pla}^*)^2}{2}\, {\Phi_*}^\dagger {\Phi_*}
   \,e^{\beta [\varphi_*^2-\varphi_{*\,\rm ini}^2]}
  \,.
\end{eqnarray}

In this frame, it is the exponential dependence of the conformal factor on $\varphi_{*\,\rm ini}^2$, what
allows to write $M_{\pla}^*$ as the unique scale.

The agreement with BBN constraints forces to an effective attraction during the radiation dominated
Universe. In this stage, it is easier to find the approximated evolution in the EF. In fact, since pure radiation
does not interact with the scalar degree of freedom at first order, it is a good approximation to assume that
the equations of motion for radiation are standard. On the other hand, the EF is defined by the fact
that the gravitational equations for the metric have their standard form. For these reasons, inside the EF,
the evolution deviates minimally from the standard case.

As it has been done in previous works for the same model (with $V_*(\varphi_*)=0$), it is convenient to use
the e-folds of expansion $p$ in the EF for studying the cosmological evolution (not to be confused with the
pressure $P$). Denoting by $R_{*,{\rm ini}}$ the value of the scale factor at the initial time
(defined by $T_*=M_{\pla}^*$), the number of e-folds is $p \equiv {\rm ln }({R_*}/R_{*,{\rm ini}})$,
which implies:
\begin{equation}
\frac{d}{d p} = \frac{1}{H_*} \, \frac{d}{d t_*} = \frac{A}{H_*} \, \frac{d }{d t}\,,
\label{chain}
\end{equation}
where prime shall denote differentiation wrt $p \,$. $R_*$ grows monotonically with time, since $H_*$ is governed by the standard Friedmann equation and it is always positive. On the contrary, the Hubble rate $H$ in the JF is related to that of the EF by
\begin{equation}
H = \frac{H_*}{A} \left( 1 + \alpha_* \, \varphi_*' \right)\,,
\label{hubbles}
\end{equation}
with $\alpha_*=d\,\ln[A(\varphi_*)]/d\,\varphi_*=\beta\,\varphi_*$. $H$ can become negative leading to a contracting universe in the JF. In a similar way, the evolution of the Jordan Temperature of the relativistic thermal bath is given by \cite{copu}:
\begin{equation}
\left( {\rm ln } \, T \right)' = - \frac{1+ \alpha_* \, \varphi_*'}{\cal I}
\simeq -(1+ \alpha_* \, \varphi_*')\,,
\label{finaleqT}
\end{equation}
since ${\cal I}-1 \lta 10^{-3}$ \cite{copu}. On the other hand, the equation for the field is given by:
\begin{equation}
\varphi_*'' = - \frac{3 - \varphi_*^{' 2}}{2} \left[ S_1(\varphi_*,\rho_*,P_*)\, \varphi_*' + S_2(\varphi_*,\rho_*,P_*) \right]\,,
\nonumber\\
\label{finaleqp}
\end{equation}
where the functions $S_1$ and $S_2$ depend on the density and pressure of the thermal bath and on $\phi_*$ trough the potential $V(\varphi_*)$ and the conformal factor $A(\varphi_*)$. In particular:
\begin{equation}
S_1(\varphi_*,\rho_*,P_*)= \frac{\rho_*-P_*+2\hat{V}(\varphi_*)}{\rho_*+\hat{V}(\varphi_*)}
\,,
\label{s1}
\end{equation}
\begin{equation}
S_2(\varphi_*,\rho_*,P_*)=
\frac{\alpha_*(\rho_*-3P_*)+\frac{d\,\hat{V}(\varphi_*)}{d\,\varphi_*}}
{\rho_*+\hat{V}(\varphi_*)}\,.
\label{s2}
\end{equation}
We have computed the evolution by assuming an ideal relativistic thermal bath: $\rho= g_f T^4 \pi^2/30$ and
$P= g_f T^4 \pi^2/90$ except for the source term: $\rho-3P$. This term is the dominant source for the evolution
of the scalar mode in the radiation dominated regimen as it has been discussed in \cite{copu}. In our case, we
have included a potential that is only important in a very first stage, when the scales unify; and in the very late
Universe, when the other energy content has been diluted and the potential behaves as a cosmological constant.
In particular, for the standard model content: $g_f\simeq 427/4$ and $\rho-3P\simeq 140\, \rho\, \alpha_3^2(T)/(61\, \pi^2)$
\cite{Kapusta:1989tk,copu}, where we are just keeping the leading modification to the non-interacting bath
coming from QCD (read \cite{copu} for a detailed discussion).

In the early evolution, the weak symmetry is not broken since $T$ is bigger than the critical temperature:
\begin{equation}
T_c^2
=
\frac{8(\varsigma\,\mu)^2}
{4\lambda+3g^2+3g'^2}
=
\frac{(348\; \gev)^2}
{1+3g^2/4\lambda+3g'^2/4\lambda}\,,
\label{Tc}
\end{equation}
where $g$, $g'$ and $\lambda$ are the standard couplings of the Weinber-Salam model \cite{Kapusta:1989tk}. In any case, it is
interesting to note that $T_c$ is very close to the Planck scale at the initial time. This is one
of the interest of this analysis, to show
that the different scales can be dynamically produced due to the evolution of the conformal factor. At this very initial time, gravitational quantum
corrections are expected to be important, and the validity of the equations is not justified
for the region where $T^* \gta M_{\pla}^*$.

\section{Discussion and Signatures}

\begin{figure}[bt]
\begin{center}
\resizebox{8.0cm}{!} {\includegraphics{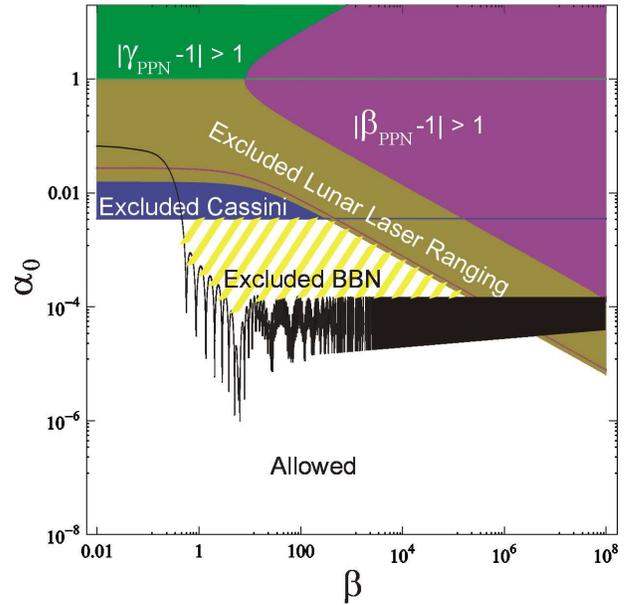}}
\caption{
Constraints in terms of the present coupling $\alpha_0\equiv \beta \varphi_*^0$
and the $\beta$ parameter. In addition to the exclusion regions from
the different data mentioned in the text, it is shown the area beyond
 the perturbative regimen of PPNs.} 
 \label{Constraints}
\end{center}
\end{figure}

The present constraints and observational signatures of this model are resumed in Fig. \ref{Constraints}.
On the one hand, the modification of the standard expansion given by Eq.~(\ref{hubbles}) originates potential deviations
in standard nucleosynthesis. The uncertainties on these observations are typically of order $10\%$ and
it roughly implies: $\beta|\varphi_*|_{nuc}^2 \lta 0.2$  and $|\beta\varphi_*\varphi_*'|_{nuc} \lta 0.1$, although
both constraints are degenerated (read Ref. \cite{couv} for a detailed analysis in the same model used
in this work but with negligible vacuum energy).

The other important signature of these models is their present modifications to GR~\cite{Will:2005va}.
The bounds are most conveniently imposed on the so called post-Newtonian parameters (PPN), which in the context
of STTs, can in turn be related to the parameter of the model: $\beta$, and the present value of the scalar: $\varphi_*^0$. 
Limits from the Very Long Baseline Interferometer \cite{Shapiro:2004zz} and radio links with the Cassini spacecraft
\cite{Bertotti:2003rm} enforce $(\beta\varphi_*^0)^2 \lta 10^{-5} \,$. The perihelion shift of Mercury \cite{mercurybound}
and the Lunar Laser Ranging experiment  \cite{Williams:1995nq} instead constrain the combination
$(\beta + 16) (\beta\varphi_*^0)^2 \lta 10^{-3}$. Therefore, $\beta$ can be large enough
in order to have an effective attraction, and reduce the present value of $\varphi_*$. Indeed, it is easy to estimate
the decrease of $\alpha$ during matter domination \cite{dnord} since the linear approximation leads to a simple damped
oscillatory solution. Just for $\beta \sim 3/8$, $\varphi_*$ decreases by a factor of about two orders of magnitude during
the matter dominated era~\cite{copu}.

Other possible signatures in gravitational lensing and the cosmic microwave background (CMB) have been studied \cite{Riaz,Sabino},
although they are less promising. In the latter, the main modification is also related to a difference in the expansion rate, which
affects the angular scale of the anisotropies. On the other hand, the decoupling time might be significantly more
perturbed and displace the peaks towards higher multipoles for faster expansions. In any case, due to the well known degeneracy of the
CMB spectrum with respect to cosmological parameters, present measurements of the peak locations
\cite{CMB} do not translate straightforwardly into a bound on the conformal factor $A$, but on a general reanalysis of data.
In any case, once the BBN bound on $A$ has been imposed, the resulting values for $A$ at photon decoupling are so close to unity as to give shifts in the peak multipoles smaller than the experimental error. Thus, the CMB spectrum does not provide significant bounds to the model under discussion.

In conclusion, we have introduced the concept of {\it Vacuum Frame} as an alternative framework to analyze the DE problem. We have illustrated such a concept by studying the pheonomenology of a simple STT, which is able to generate the large hierarchies among the Planck, electroweak and cosmological constant scale in a dynamical way. 

\begin{acknowledgments}
This work was partially supported by the MICINN (Spain) project PID2019-107394GB-I00 (AEI/FEDER, UE).
I acknowledge support by Institut Pascal at Université Paris-Saclay during the Paris-Saclay Astroparticle Symposium 2021, with the support of the P2IO Laboratory of Excellence (program “Investissements d’avenir” ANR-11-IDEX-0003-01 Paris-Saclay and ANR-10-LABX-0038), the P2I axis of the Graduate School Physics of Université Paris-Saclay, as well as IJCLab, CEA, IPhT, APPEC, the IN2P3 master projet UCMN and EuCAPT.
This research was supported by the Munich Institute for Astro- and Particle Physics (MIAPP) which is funded by the Deutsche Forschungsgemeinschaft (DFG, German Research Foundation) under Germany´s Excellence Strategy – EXC-2094 – 390783311.
\end{acknowledgments}

\vspace{.2 cm}


\begin{references}

\bibitem{stgen}
 P. Jordan, Nature (London) {\bf 164}, 637 (1956);\\
 M. Fierz, Helv. Phys. Acta {\bf29}, 128 (1956);\\
 C. Brans and R. Dicke, Phys. Rev. D {\bf124}, 925 (1961);\\
 P.G. Bergmann, Int. J. Theor. Phys. {\bf1}, 25 (1968);\\
 K. Nordtvedt, Astrophys. J. {\bf 161}, 1059 (1970);\\
 R. Wagoner, Phys. Rev. D{\bf 1}, 3209 (1970);\\
 J.~A.~R.~Cembranos, Phys. Rev. D \textbf{73}, 064029 (2006);\\
 J.~A.~R.~Cembranos, Phys. Rev. Lett. \textbf{102}, 141301 (2009).

\bibitem{Kalara:1990ar}
  S.~Kalara, N.~Kaloper and K.~A.~Olive,
  Nucl.\ Phys.\  B {\bf 341}, 252 (1990).

\bibitem{polchy}
 J. Polchinsky, {\it String theory}
 (Cambridge University Press, 1998).

\bibitem{dnord}
   T.~Damour and K.~Nordtvedt,
  Phys.\ Rev.\ Lett.\  {\bf 70}, 2217 (1993);
%
  Phys.\ Rev.\  D {\bf 48}, 3436 (1993).

\bibitem{dpol}
  T.~Damour and A.~M.~Polyakov,
  Nucl.\ Phys.\  B {\bf 423}, 532 (1994).

\bibitem{couv}
  A.~Coc, K.~A.~Olive, J.~P.~Uzan and E.~Vangioni,
  Phys.\ Rev.\  D {\bf 73}, 083525 (2006);\\
  A.~Coc, K.~A.~Olive, J.~P.~Uzan and E.~Vangioni,
  Phys. Rev. D \textbf{79}, 103512 (2009).

\bibitem{copu}
  J.~A.~R.~Cembranos, K.~A.~Olive, M.~Peloso and J.~P.~Uzan,
	JCAP \textbf{07}, 025 (2009).

\bibitem{Kapusta:1989tk}
  J.~I.~Kapusta,
  "Finite Temperature Field Theory", Cambridge (1989).

\bibitem{Will:2005va}
  C.~M.~Will,
  Living Rev.\ Rel.\  {\bf 9}, 3 (2005).

\bibitem{Shapiro:2004zz}
  S.~S.~Shapiro, J.~L.~Davis, D.~E.~Lebach and J.~S.~Gregory,
  Phys.\ Rev.\ Lett.\  {\bf 92}, 121101 (2004).

\bibitem{Bertotti:2003rm}
  B.~Bertotti, L.~Iess and P.~Tortora,
  Nature {\bf 425}, 374 (2003).

\bibitem{mercurybound}
 I.I. Shapiro, in {\it General Relativity and Gravitation 12},
 N. Ashby {\it et al.} Eds. (Cambridge University Press, 1990),
 pp. 313.

\bibitem{Williams:1995nq}
  J.~G.~Williams, X.~X.~Newhall and J.~O.~Dickey,
  Phys.\ Rev.\  D {\bf 53}, 6730 (1996).

\bibitem{Riaz} A.~Riazuelo and J.~P.~Uzan,
Phys.\ Rev.\ D {\bf 66}, 023525 (2002).

\bibitem{Sabino}
X.~l.~Chen and M.~Kamionkowski,
Phys.\ Rev.\ D {\bf 60}, 104036 (1999);\\
F.~Perrotta, C.~Baccigalupi and S.~Matarrese,
Phys.\ Rev.\ D {\bf 61}, 023507 (2000);\\
R.~Nagata, T.~Chiba and N.~Sugiyama, Phys.\ Rev.\ D {\bf 66}, 103510 (2002);\\
R.~Nagata, T.~Chiba and N.~Sugiyama, Phys.\ Rev.\ D {\bf 69}, 083512 (2004).

\bibitem{CMB} P.~de Bernardis {\it et al.},
Astrophys.\ J.\  {\bf 564}, 559 (2002); R.~Stompor {\it et al.},
Astrophys.\ J.\  {\bf 561}, L7 (2001);\\
C.~L.~Bennett {\it et al.},
Astrophys.\ J.\ Suppl.\  {\bf 148}, 1 (2003).

\end{references}
\end{document}